\documentclass[aps,amssymb,floatfix,prd,amsmath,preprintnumbers]{revtex4}
\usepackage{xcolor}
\usepackage{graphicx}  
\usepackage{float}
\usepackage{dcolumn}   
\usepackage{bm}
\begin{document}
\title{\bf Stability analysis of two-fluid dark energy models}

\author{B. Mishra\footnote{Department of Mathematics, Birla Institute of Technology and Science-Pilani, Hyderabad Campus, Hyderabad-500078, India, E-mail: bivu@hyderabad.bits-pilani.ac.in}, Fakhereh Md. Esmaeili \footnote{School of Physics, University of Hyderabad, Hyderabad-500046, India, E-mail:astrosat92@gmail.com}, Pratik P. Ray,\footnote{Department of Mathematics (SSL), Vellore Institute of Technology-Andhra Pradesh University, Andhra Pradesh - 522237, India}, S. K. Tripathy\footnote{Department of Physics, Indira Gandhi Institute of Technology, Sarang, Dhenkanal, Odisha-759146, India, E-mail:tripathy\_ sunil@rediffmail.com}}
\affiliation{ }
\begin{abstract}
In this paper, we have studied the stability of the cosmological models with dark energy and combination of matter fields and dark energy in an anisotropic space time. The pressure anisotropy along the spatial directions are derived and its stability in each direction has been examined. The four models presented here, show its stability on certain spatial direction. The role of matter field on the stability analysis has been obtained. The positive and negative value of cosmic string completely changed the stability behaviour of the model. The presence of a magnetic field disturbs the stability aspects of the models at least in an early epoch. 
\end{abstract}
\maketitle

\section{Introduction}

In theoretical cosmology and astrophysics problems, the field equations obtained are highly non-linear. Several assumptions are normally considered to obtain the exact solution. It has been difficult to assesses the degree of the generality of the solution because of the limitations in the consideration. However, the asymptotic behaviour provides relevant features to compare with physical results. Therefore, in order to further strengthen the results of the cosmological models, researcher analyzing the qualitative properties of the field equations. The appropriate approach for analyzing this, is the stability analysis \cite{Wainright97,Charters01}. Prior to this, in order to show the stability in relativity, the Liapunov’s method has been used \cite{Heusler91,Sudarsky95}. With the cosmological tests, Farajollahi et al. \cite{Farajollahi11a}  have shown that the universe starts from an unstable state and with the evolution progress, it became stable near steady state. Also, Farajollahi and Salehi \cite{Farajollahi11b} have fitted the stability and model parameters with the observational data to obtain the accelerating model developed with Brans Dicke cosmology.  Nozari et al. \cite{Nozari12}, found its critical points from the models in phase space and studied its stability solution correspond to the accelerated phase of the expansion of the Universe. Similarly, Mishra and Chakraborty \cite{Mishra19c} with stability analysis have shown that the initial position of the universe is close to an unstable critical point and ends close to a stable critical point.\\

In recent times, several theoretical ideas have been explored to understand the origin of dark energy (DE), which has occupied a major portion in the mass energy budget of the universe. It is believed that the presence of dark energy is responsible for the present universe to expand and accelerate. Several theoretical dark energy models have been presented in different context. Li et al. \cite{Li10} have compared the dark energy cosmological models in an isotropic space time based on the observational data. Yoo and Watanabe \cite{Yoo12} with some observational evidences revealed that the theoretical models for dark energy can be studied through inhomogeneous model, cosmological  constant, modified gravity model and modified matter model. Mishra and Tripathy \cite{Mishra15} have shown the anisotropic behaviour of the cosmological model constructed with an anisotropic metric. Xu and Zhang \cite{Xu16} compared ten DE models according to their capabilities to fit the observational data and have shown that $\Lambda$CDM model is best model among them. Mishra et al. \cite{Mishra17} derived the cosmological models using general form of scale factors and shown the accelerating universe using the representative values of the model parameters.  Khurshudyan and Khurshudyan \cite{Khurshudyan18} have  investigated several cosmological models with a non-linear forms of interaction between DE and cold dark matter.  Farnes \cite{Farnes18} has proposed a cosmological model that predicts the observed distribution of dark matter in galaxies.  Mishra et al. \cite{Mishra18a} presented the cosmological models in a two fluid scenario and have shown the role of matter field in the accelerating universe. Tiwari et al. \cite{Tiwari18} presented the DE cosmological models in an anisotropic space time. In order to constraint the conformal, disformal and mixed interaction between DE and dark matter Bruck and Mifsud \cite{Bruck18} have performed a global analysis with the help of cosmological data sets. Mishra et al. \cite{Mishra18b} have shown the role of cosmic string in the study of expanding universe. \\

Pal and Chakraborty \cite{Pal19} have suggested three-fluid cosmological model  consisting of baryonic matter, DE and non-interacting dark matter in an isotropic background. Tawfik and Dahab \cite{Tawfik19} have reviewed the DE models based on the holographic principle and studied the stability based on the cosmological perturbation. Mishra et al. \cite{Mishra19a} have shown the effect of bulk viscous fluid in the cosmic expansion of the universe in a two fluid scenario. Odintsov and Oikonomou \cite{Odintsov19} have investigated the singularity structure of the phase space of an exponential quintessence model. Sadri and  Khurshudyan \cite{Sadri19} have shown that the deceleration parameter and the equation of state parameter depict an accelerated universe interacting new holographic DE models. Ray et al. \cite{Ray19} and Mishra et al. \cite{Mishra19d} have analysed DE energy with electromagnetic field along $x$ and $z$- axes respectively. They have shown that the electromagnetic field has a greater impact as compared to other matter sources in the late time cosmic acceleration of the universe. With the interaction of dark matter and DE, Cheng et al. \cite{Cheng20} have studied the impact of the interaction on the accelerating universe. Valentino et al. \cite{Valentino20} examined the interactions between dark matter and DE in light of the latest cosmological observations with a focus on a specific model with coupling proportional to the DE density. Referring to the recent $H(z)$ and Pantheon data, Goswami et al. \cite{Goswami20} have constrained the cosmological parameters in Bianchi type V space time. Kritpetch et al. \cite{Kritpetch20} have investigated the non-minimal derivative coupling in holographic DE model. Amirhashchi and Amirhashchi \cite{Amirhashchi20} have constrained the cosmological parameters of Bianchi I model with $H(z)$ and type Ia supernovae data.  \\

In this paper, we are interested to study the stability analysis of the cosmological models framed in Bianchi type V space time with the combinations of matter source and DE. Then a comparative study would be done on the stability behaviours of the models with various celestial matter present in our universe. The motivation behind choosing the two fluid scenario is that the matter field in addition to the DE plays a significant role in the study of accelerating universe. We are interested here to examine its stability corresponding to the spatial directions. In section II, the mathematical frame up has been done with the derivation of its physical parameters. Stability analysis for each case has been done in section III. The results and conclusion are given in section IV.

\section{Mathematical set up with Physical Parameters}

In order to frame the cosmological model of DE, we have considered the spatial homogeneous anisotropic Bianchi V space time in the form 

\begin{equation} \label{eq:1}
 ds^{2}= dt^{2}-a(t)^{2} dx^{2}-e^{2\alpha x}[b(t)^{2} dy^{2}+c(t)^{2} dz^{2}]
\end{equation}

where the metric potentials  $a$,$b$ and $c$ are function of cosmic time $t$ only and $\alpha$ is constant. Since the metric potentials are different in different directions, the space time provides a source of anisotropy. Let us consider that General Relativity (GR) is well defined at cosmic scales. Hence, the Einstein's field equations can be defined as 

\begin{equation} \label{eq:2}
G_{ij}\cong R_{ij}-\frac{1}{2}Rg_{ij}=\kappa T_{ij},
\end{equation}

where, $R_{ij}$, $R$, $G_{ij}$ and $T_{ij}$ respectively denote the Ricci tensor, Ricci scalar, Einstein tensor and the effective energy momentum tensor (EMT). Further $T_{ij}=T^{M}_{ij}+T^{D}_{ij}$, where $T^{M}_{ij}$ and $T^{D}_{ij}$ respectively represent the EMT of matter and DE respectively. We assume $\kappa = - \frac{8 \pi G}{c^{4}}$ and $8 \pi G=c =1$ for the the physical convenience. So, the field eqns. \eqref{eq:2} with the effective EMT and in the frame GR can be defined as  

\begin{eqnarray}
\frac{\ddot{b}}{b}+\frac{\ddot{c}}{c}+\frac{\dot{b}\dot{c}}{bc}-\frac{\alpha^{2}}{a^{2}}&=&\frac{1}{a^2}\left[T^{M}_{11}+T^{D}_{11}\right] \label{eq:3} \\
\frac{\ddot{a}}{a}+\frac{\ddot{c}}{c}+\frac{\dot{a}\dot{c}}{ac}-\frac{\alpha^{2}}{a^{2}}&=&\frac{1}{b^2e^{2\alpha x}}\left[T^{M}_{22}+T^{D}_{22}\right] \label{eq:4} \\
\frac{\ddot{a}}{a}+\frac{\ddot{b}}{b}+\frac{\dot{a}\dot{b}}{ab}-\frac{\alpha^{2}}{a^{2}}&=&\frac{1}{c^2e^{2\alpha x}}\left[T^{M}_{33}+T^{D}_{33}\right] \label{eq:5} \\
\frac{\dot{a}\dot{b}}{ab}+\frac{\dot{b}\dot{c}}{bc}+\frac{\dot{c}\dot{a}}{ca}-\frac{3\alpha^{2}}{a^{2}}&=& \left[T^{M}_{44}+T^{D}_{44}\right]   \label{eq:6} \\
2\dfrac{\dot{a}}{a}-\dfrac{\dot{b}}{b}-\dfrac{\dot{c}}{c}&=&0  \label{eq:7}
\end{eqnarray}

Eqn. \eqref{eq:7} gives $a=k_1bc$, where $k_1$ is the integrating constant and is considered to be unity. An over dot represents the derivative with respect to the cosmic time $t$.\\

We can transform the set of field eqns.\eqref{eq:3}-\eqref{eq:7} in to the form of Hubble parameter with the background $H_x=\frac{\dot{a}}{a}$, $H_y=\frac{\dot{b}}{b}$, $H_z=\frac{\dot{c}}{c}$. The mean Hubble parameter can be obtained as $H=\frac{\dot{\mathcal{R}}}{\mathcal{R}}=\frac{1}{3}(H_x+H_y+H_z)$, where $\mathcal{R}$ is the scale factor. The shear scalar $\sigma^2\left[=\sigma_{ij}\sigma^{ij}=\frac{1}{2} \left(\Sigma H^2_i-\frac{\theta^2}{3}\right)\right]$ usually considered to be proportional to the scalar expansion $\theta(=3H)$ \cite{Shamir10, Tripathy2015, Mishra15, Tripathy2015a, Tripathy2021}. This leads to an anisotropy relation between the metric potentials $b$ and $c$ in the form $b=c^m$. Now, the relationship between the scale factor and the metric potentials can be established as $a=\mathcal{R}$, $b=\mathcal{R}^{\frac{2m}{m+1}}$ , $c=\mathcal{R}^{\frac{2}{m+1}}$. Also the Hubble parameter and directional Hubble parameters can be related as $H_x=H$, $H_{y}=\left(\frac{2m}{m+1}\right)H$ and $H_{z}=\left(\frac{2}{m+1}\right)H$. It can be noted that the rate of expansion along $x$-axis is same as the mean Hubble parameter. In the energy momentum tensor of DE ($T^{D}_{ij}$), we have also incorporated some degree of anisotropy in the DE pressure, which can be represented as
\begin{eqnarray} \nonumber
T^D_{ij}&=& diag\left[\rho_D, -p_{Dx},-p_{Dy},-p_{Dz}\right]\\ \nonumber
&=& diag\left[1, -\omega_{Dx}, -\omega_{Dy}, -\omega_{Dz} \right] \rho_D\\ 
&=& diag\left[1, -(\omega_D+\delta), -(\omega_D+\gamma), -(\omega_D+\eta)\right]\rho_D  \label{eq:8}
\end{eqnarray}
where $p_{Dx}$, $p_{Dy}$ and $p_{Dz}$ respectively denotes the DE pressure along $x$,$y$ and $z$-axes. It is worth to mention here that, $\Lambda$CDM model with cosmological constant does not require any dark energy form to explain the late time  Cosmic acceleration. The equation of state parameter corresponding to the concordance is $\Lambda$CDM model. Usually, the vacuum energy density in such model is associated with the cosmological constant which provides the anitgravity affect and drives the cosmic acceleration. However, the fine tunning problem in theoretical physics and coincidence problem in cosmology require that the vacuum energy density should roll down from a high value to an observationally acceptable small but positive value. In view of this different dynamical DE models have been proposed in literature. Usually in these dynamical DE models, some scalar fields such as quintessence field, phantom field, k-essence, tachyons etc. are used to understand the dynamics of late time cosmic speed up phenomena. These scalar fields are then confronted with observations to obtain viable cosmological models. In the present work, we have considered a contribution of dark energy in the matter side of the field equation in the frame work of General Relativity. The DE is assumed to be dynamical. Also, we allow a pressure anisotropy due to the dark energy along different spatial directions.  Accordingly we have defined the energy momentum tensor for the DE contribution in eq.\eqref{eq:8} \cite{MST2015, TMS2017}.

The field eqns. \eqref{eq:3}-\eqref{eq:7} can be expressed in the form of Hubble parameter as 

\begin{eqnarray}
2\dot{H}+4\left(\frac{m^2+m+1}{m^2+2m+1}\right)H^2-\frac{\alpha^2}{a^2}&=&\frac{1}{a^2}\left[T^{M}_{11}\right]-(\omega_D+\delta)\rho_D, \label{eq:9} \\
\left(\frac{m+3}{m+1}\right)\dot{H}+\left(\frac{m^2+4m+7}{m^2+2m+1}\right)H^2-\frac{\alpha^{2}}{a^{2}}&=&\frac{1}{b^2e^{2\alpha x}}\left[T^{M}_{22}\right]-(\omega_D+\gamma)\rho_D, \label{eq:10} \\
\left(\frac{3m+1}{m+1}\right)\dot{H}+\left(\frac{7m^2+4m+1}{m^2+2m+1}\right)H^2-\frac{\alpha^{2}}{a^{2}}&=&\frac{1}{c^2e^{2\alpha x}}\left[T^{M}_{33}\right]-(\omega_D+\eta)\rho_D, \label{eq:11} \\
2\left(\frac{m^2+4m+1}{m^2+2m+1}\right)H^2-\frac{3\alpha^{2}}{a^{2}}&=& \left[T^{M}_{44}\right] +\rho_{D}.  \label{eq:12}
\end{eqnarray}

\section{Stability Analysis}
In this section, we are intending to undertake the stability analysis of the anisotropic DE model with various matter source using the hybrid scale factor described as $\mathcal{R}=e^{\mu t}t^{\nu}$; so as the Hubble parameter, $H=\frac{\dot{\mathcal{R}}}{\mathcal{R}}=\mu+\frac{\nu}{t}$. This reduces to power law when $\mu=0$ and to exponential law when $\nu=0$. The stability analysis would be performed for the cosmological models with DE only \cite{Mishra15} and models with two fluid scenario such as, DE with (i) viscous fluid \cite{Mishra19a}, (ii) cosmic string \cite{Mishra18b} and  (iii) electromagnetic field \cite{Ray19}. The stability of the model can be analysed by considering the mechanical stability of the cosmic fluid. This is usually tested through the calculation of the adiabatic speed of sound through the cosmic fluid, $C_{s}^{2}= \dfrac{dp}{d \rho_D}$ \cite{Xu12,Xu13,Balbi07}, measured in the unit of the square of the speed of light in vacuum. The stability condition occurs for a positive squared sound speed $C_{s}^{2}>0$. Also the cosmic fluid should not have a supraluminous behaviour with $C_{s}^{2}>1$. Xu has obtained that $C_{s}^{2}=0.00155^{+0.000319}_{-0.00155}$ \cite{Xu13} and Xu et al. as $C_{s}^{2}=0.000487^{+0.000117}_{-0.000487}$ \cite{Xu12}. Liao et al. have obtained the adiabatic squared sound speed as $C_{s}^{2}=0.00242^{+0.00787}_{-0.00775}$ \cite{Liao2012}.  It is to note here that, we have presented a mathematical formalism  that involves directional anisotropic pressures along different orthogonal spatial coordinates. In view of this, we wish to calculate the adiabatic speed of sound in respective directions and analyse the respective directional functionals $(C_{sx}^{2}= \dfrac{d\delta}{d\rho_D}$, $ C_{sy}^{2}= \dfrac{d\gamma}{d \rho_D},$ $ C_{sz}^{2}= \dfrac{d\eta}{d \rho_D} )$ to test the stability of the model. Here, the additional suffixes $x,y$ and $z$ denote the functional in the respective axis. \\

\subsection{Model I}
We have considered, the energy momentum tensor as DE fluid only, then we the field eqns. \eqref{eq:9}-\eqref{eq:12} can be reduced to \cite{Mishra15},
\begin{eqnarray}
2\dot{H}+4\left(\frac{m^2+m+1}{m^2+2m+1}\right)H^2-\frac{\alpha^2}{\mathcal{R}^2}&=&-(\omega_D+\delta)\rho_D \label{eq:14} \\
\left(\frac{m+3}{m+1}\right)\dot{H}+\left(\frac{m^2+4m+7}{m^2+2m+1}\right)H^2-\frac{\alpha^{2}}{\mathcal{R}^{2}}&=&-(\omega_D+\gamma)\rho_D \label{eq:15} \\
\left(\frac{3m+1}{m+1}\right)\dot{H}+\left(\frac{7m^2+4m+1}{m^2+2m+1}\right)H^2-\frac{\alpha^{2}}{\mathcal{R}^{2}}&=&-(\omega_D+\eta)\rho_D \label{eq:16} \\
2\left(\frac{m^2+4m+1}{m^2+2m+1}\right)H^2-\frac{3\alpha^{2}}{\mathcal{R}^{2}}&=& \rho_{D}  \label{eq:17}
\end{eqnarray}

With the algebraic manipulation, from eqns. \eqref{eq:14}-\eqref{eq:16}, we can obtain

\begin{eqnarray}  
\delta &=& -\frac{2}{3\rho_D} \left( \frac{m^2-2m+1}{m^2+2m+1} \right) F(R) \label{eq:18} \\ 
\gamma &=& \frac{1}{3\rho_D} \left( \frac{m^2+4m-5}{m^2+2m+1} \right) F(R) \label{eq:19} \\ 
\eta &=& -\frac{1}{3\rho_D} \left( \frac{5m^2-4m-1}{m^2+2m+1} \right) F(R) \label{eq:20}
\end{eqnarray}

where, $F(R)=\frac{\ddot{\mathcal{R}}}{\mathcal{R}}+2\frac{\dot{\mathcal{R}}^2}{\mathcal{R}^2}$ and the DE density $\rho_D$ for the hybrid scale factor can be derived as,

\begin{equation}\label{DEDP}
\rho_D=2\left(\frac{m^2+4m+1}{m^2+2m+1}\right)\left(\mu+\frac{\nu}{t}\right)^2-3\left(\frac{\alpha}{e^{\mu t}t^{\nu}}\right)^2.
\end{equation}

The present DE model is based on the work of Mishra and Tripathy \cite{Mishra15} and therefore, we restrict the parameters used in here as the constraints obtained in Ref. \cite{Mishra15}. The nature of the hybrid scale factor is to display a cosmic transit behaviour with an early deceleration and late time acceleration simulated by a signature flipping behaviour of the deceleration parameter. In order to show a transit from a decelerating behaviour to an accelerated one, we require that $\nu$ should remain in the range $0 <\nu <1$. Moreover, the functional form $F(R)$ developed in the formulation suggests a tighter range $[0,\frac{1}{3}]$ for $\nu$, so that at an early epoch we may have a decelerated Universe. At the same time the parameter $\mu>0$ is considered as a free parameter. It is also worthy to mention here that the Universe is almost isotropic and there have been many attempts in recent times to pin down the cosmic anisotropy, if any, present in the Universe \cite{Bunn1996,Nilsson1999, Saadeh2016, Deng2018, Deng2018a, Chang2019}. A tighter constraint on the cosmic anisotropy has been obtained from these results. Corresponding to an average anisotropy $4.4439 \times 10^{-9}$, we have fixed the value of the anisotropy parameter from our Bianchi V model as $m = 1.0001633$ \cite{Mishra15}. This corresponds to $\left(\frac{\sigma}{H}\right)_0 \simeq 8.164 \times 10^{-5}$. We shall adhere to this value of the anisotropic parameter in all the following models to examine the stability aspect. The metric exponent parameter $\alpha$ is chosen to be a small quantity so that the model closely resembles an FRW model for same values of the directional scale factors. In the present work, we chose $\alpha=0.05$, so that energy density becomes positive and we do not deny the possibility of other choices of this parameter. \\

Now, using eqns. \eqref{eq:18}-\eqref{DEDP}, the stability function $C_{sx}^2$, $C_{sy}^2$ and $C_{sz}^2$ can be obtained and its graphical behaviour has been shown in FIG. 1. 
\begin{figure}[h!]
\centering
\includegraphics[width=0.8\textwidth]{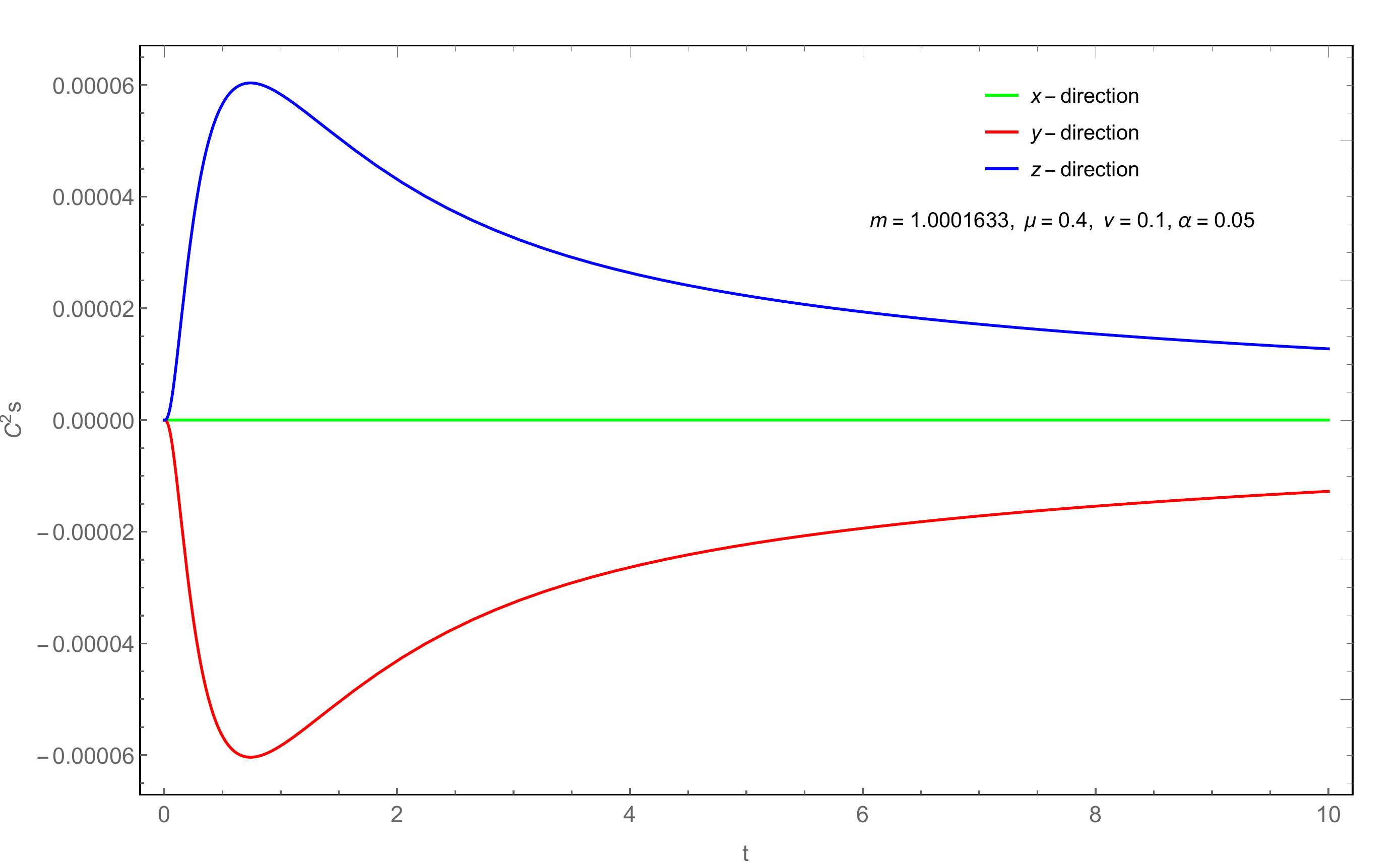}
\caption{Stability analysis for DE model I. The unit of the speed of sound $C_s$ is considered as the speed of light in vacuum. The time scale is chosen in the unit of the present age of the Universe i.e at $t=1$, we have the present epoch. } 
\end{figure}

From FIG. 1, it may be observed that the model, is stable in the $z$-direction whereas along the $y$ direction, the model appears to be unstable. In fact, the pressure along the $y$ direction is less compared to the isotropic pressure which provide a negative value to the directional pressure component. This leads to a negative value for $C_{sy}^2$. However, if we consider only the magnitude of the departure of the anisotropic DE pressure from the isotropic value, we may get a positive quantity for $C_{sy}^2$. The mechanical stability condition is just satisfied along the $x$-direction. This behaviour along $x$-direction might be due to the initial consideration of the equality between the metric potential along $x$-direction and the average scale factor $\mathcal{R}$. \\

\subsection{Model II}
In this case, we have considered the energy momentum tensor as the combination of matter as viscous fluid and the DE \cite{Mishra19a}, which can be expressed as,  
\begin{equation}
T^M_{ij}=(\rho+\bar{p})u_iu_j-pg_{ij}\label{EMTV}
\end{equation}
and the energy conservation equation yields
\begin{equation} \label{ECEV}
\dot{\rho}+3(\rho+\bar{p})=0.
\end{equation}
Here $u^i$ is the four velocity vector of the fluid in a co moving coordinate system, $\bar{p}$ is total pressure including the contribution from a bulk viscous fluid. The four velocity vector satisfies  the relation $u_{i}u^{j}=1$. The field eqns. \eqref{eq:9}-\eqref{eq:12} yield
\begin{eqnarray}
2\dot{H}+4\left(\frac{m^2+m+1}{m^2+2m+1}\right)H^2-\frac{\alpha^2}{\mathcal{R}^2}&=&-\bar{p}-(\omega_D+\delta)\rho_D \label{FEV1} \\
\left(\frac{m+3}{m+1}\right)\dot{H}+\left(\frac{m^2+4m+7}{m^2+2m+1}\right)H^2-\frac{\alpha^{2}}{\mathcal{R}^{2}}&=&-\bar{p}-(\omega_D+\gamma)\rho_D \label{FEV2} \\
\left(\frac{3m+1}{m+1}\right)\dot{H}+\left(\frac{7m^2+4m+1}{m^2+2m+1}\right)H^2-\frac{\alpha^{2}}{\mathcal{R}^{2}}&=&-\bar{p}-(\omega_D+\eta)\rho_D \label{FEV3} \\
2\left(\frac{m^2+4m+1}{m^2+2m+1}\right)H^2-\frac{3\alpha^{2}}{\mathcal{R}^{2}}&=& \rho +\rho_{D}  \label{FEV4}
\end{eqnarray}

From eqns. \eqref{FEV1}-\eqref{FEV3}, we can obtain the skewness parameters as in eqns. \eqref{eq:18}-\eqref{eq:20} and the DE energy density can be calculated from eq. \eqref{FEV4} with respect to the matter energy density $\rho$. This situation is arising due to the consideration of two fluid scenario for the energy momentum tensor. The matter energy density $\rho$ can be derived from the conservation equations as,

\begin{equation}\label{ME}
\rho=\frac{\rho_0}{\left[e^{\int H.dt}\right]^{3(\epsilon+1)}}
\end{equation}
 
where, $\rho_0$ is the integrating constant and can be interpreted as the rest energy density at present time and $\epsilon$ be the viscous coefficient. Now the stability function $C_{sx}^2$, $C_{sy}^2$ and $C_{sz}^2$ can be represented graphically as in FIG. 2. 
\begin{figure}[h!]
\centering
\includegraphics[width=0.8\textwidth]{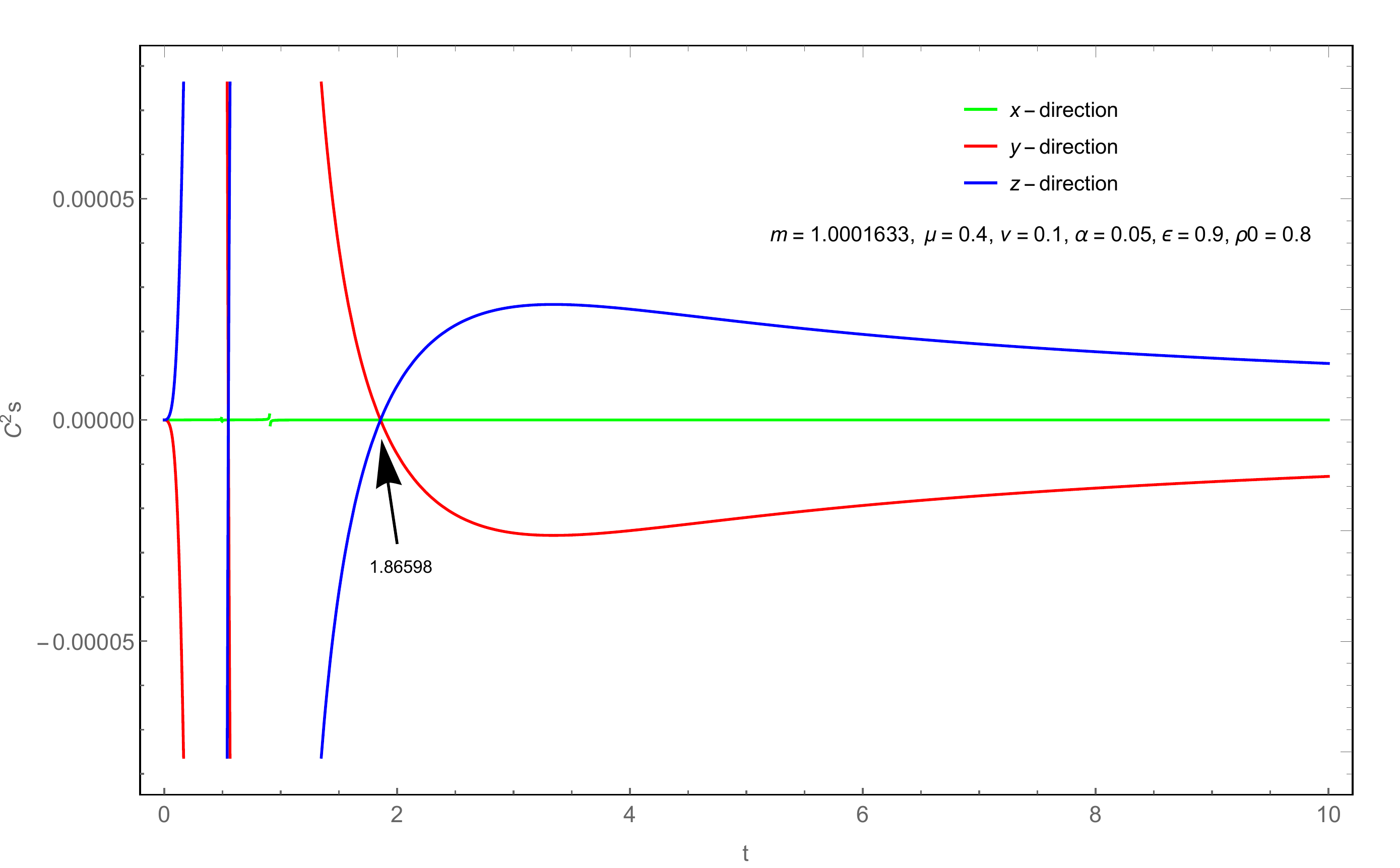}
\caption{Stability analysis for the two fluid DE scenario with a viscous fluid and DE in Model II. The unit of the speed of sound $C_s$ is considered as the speed of light in vacuum. The time scale is chosen in the unit of the present age of the Universe i.e at $t=1$, we have the present epoch.} 
\end{figure}
Here, the model shows a signature flipping behaviour along $y$ and $z$-axes, though along  $x$-axis, it still remains same as in model- I. At the initial stage, the model remains stable along y-axis and unstable along z-axis, but after the intersecting point $(1.86598)$, they changed the behaviour completely. So, most of the time the model along $z$-axis is stable and along $y$-axis, it remains unstable. This behaviour is might be due the extra amount of pressure gather through the bulk viscous fluid in the energy momentum tensor.
\subsection{Model III}
Out of different topological  stable defects, the one dimensional cosmic strings arise as a random network of line-like defects during the phase transition in the early universe with spontaneous symmetry breaking. In fact the confined regions of the false vacuum form a locus of trapped energy conceived as a self gravitating cosmic string. The energy scale at which the phase transition takes place determines the mass and dimension of a cosmic string. The mass per unit length of cosmic string is of the order of the GUT scale, $G\mu\simeq 10^{-7}$. Massive closed loops of string are believed to serve as seeds for the formation of large scale structures like galaxies and cluster of galaxies \cite{Vilenkin1981}.  While matter is accreted onto loops, the one dimensional cosmic strings oscillate violently and lose their energy by gravitational radiation thereby shrink to disappear \cite{Vilenkin1981}. This radiation bears the signature of early epoch cosmic strings and  may be detectable in experiments for gravitational waves. However, there are no observational confirmation regarding the existence of cosmic strings. Cosmic strings may be able to induce some temperature anisotropy in Cosmic Microwave Background (CMB) radiation. The mean angular power spectrum of string-induced CMB temperature anisotropies can be described by a power law. This suggests a non-vanishing string contribution to the overall CMB anisotropies may become the dominant source of fluctuations at small angular scales \cite{Fraisse2008}. There occur inconsistencies in the power spectrum of the string induced CMB temperature anisotropies. Also the radiative effects of cosmic strings are rapidly damped \cite{Gregory1989}. However, recently, Slagter has shown that, the alignment of the polarization axes of quasars in large quasar groups on Mpc scales, can be explained by cosmic string networks in the framework of General Relativity and therefore raises a possibility to detect them \cite{Slagter2017}. In the present work, we consider one dimensional cosmic strings as an additional source of cosmic anisotropy in our model and obtain the stability conditions.

With the matter field represented by one dimensional cosmic string \cite{Mishra18b}, the energy momentum tensor can be defined as
\begin{equation}
T^M_{ij}=(\rho+p)u_iu_j-pg_{ij}+\lambda x_ix_j \label{eq:21}
\end{equation}
where $x^i$ represents the string direction and $x_{i}x^{j}=-1$ (along $x$- direction). Also, $x_i$ are orthogonal to the four velocity vector i.e $x^{i}u_i=0$. In the above eqn. \eqref{eq:21}, $\lambda$ is the string tension density which may either be positive or negative.

In presence of one-dimensional cosmic string in addition the DE fluid, the field eqns. \eqref{eq:9}- \eqref{eq:12} become
\begin{eqnarray}
2\dot{H}+4\left(\frac{m^2+m+1}{m^2+2m+1}\right)H^2-\frac{\alpha^2}{\mathcal{R}^2}&=&-p-\lambda-(\omega_D+\delta)\rho_D \label{eq:22} \\
\left(\frac{m+3}{m+1}\right)\dot{H}+\left(\frac{m^2+4m+7}{m^2+2m+1}\right)H^2-\frac{\alpha^{2}}{\mathcal{R}^{2}}&=&-p-(\omega_D+\gamma)\rho_D \label{eq:23} \\
\left(\frac{3m+1}{m+1}\right)\dot{H}+\left(\frac{7m^2+4m+1}{m^2+2m+1}\right)H^2-\frac{\alpha^{2}}{\mathcal{R}^{2}}&=&-p-(\omega_D+\eta)\rho_D \label{eq:24} \\
2\left(\frac{m^2+4m+1}{m^2+2m+1}\right)H^2-\frac{3\alpha^{2}}{\mathcal{R}^{2}}&=& \rho +\rho_{D}.  \label{eq:25}
\end{eqnarray}

In eqn. \eqref{eq:22}, we assume a linear string equation of state $\lambda=3\xi\rho$ where where $\xi$ be the state parameter. For a constant equation of state parameter $\omega$, we may have $\rho=\rho_0\mathcal{R}^{-3(1+\omega+\xi)}$ where $\rho_0$ be the energy density of the matter at the present epoch. Consequently, the string tension density becomes $\lambda=3\xi\rho_0\mathcal{R}^{-3(1+\omega+\xi)}$. Eqns. \eqref{eq:22}-\eqref{eq:24} provide the skewness parameters $\gamma$ and $\eta$ as in \eqref{eq:19} and \eqref{eq:20}
respectively, however the parameter along $x$-axis can be obtained as,

\begin{eqnarray}  
\delta &=& -\frac{2}{3\rho_D}\left[ \left( \frac{m^2-2m+1}{m^2+2m+1} \right) F(R)+\lambda \right]. \label{eq:26}
\end{eqnarray}

\begin{figure}[h!]
\centering
\includegraphics[width=0.8\textwidth]{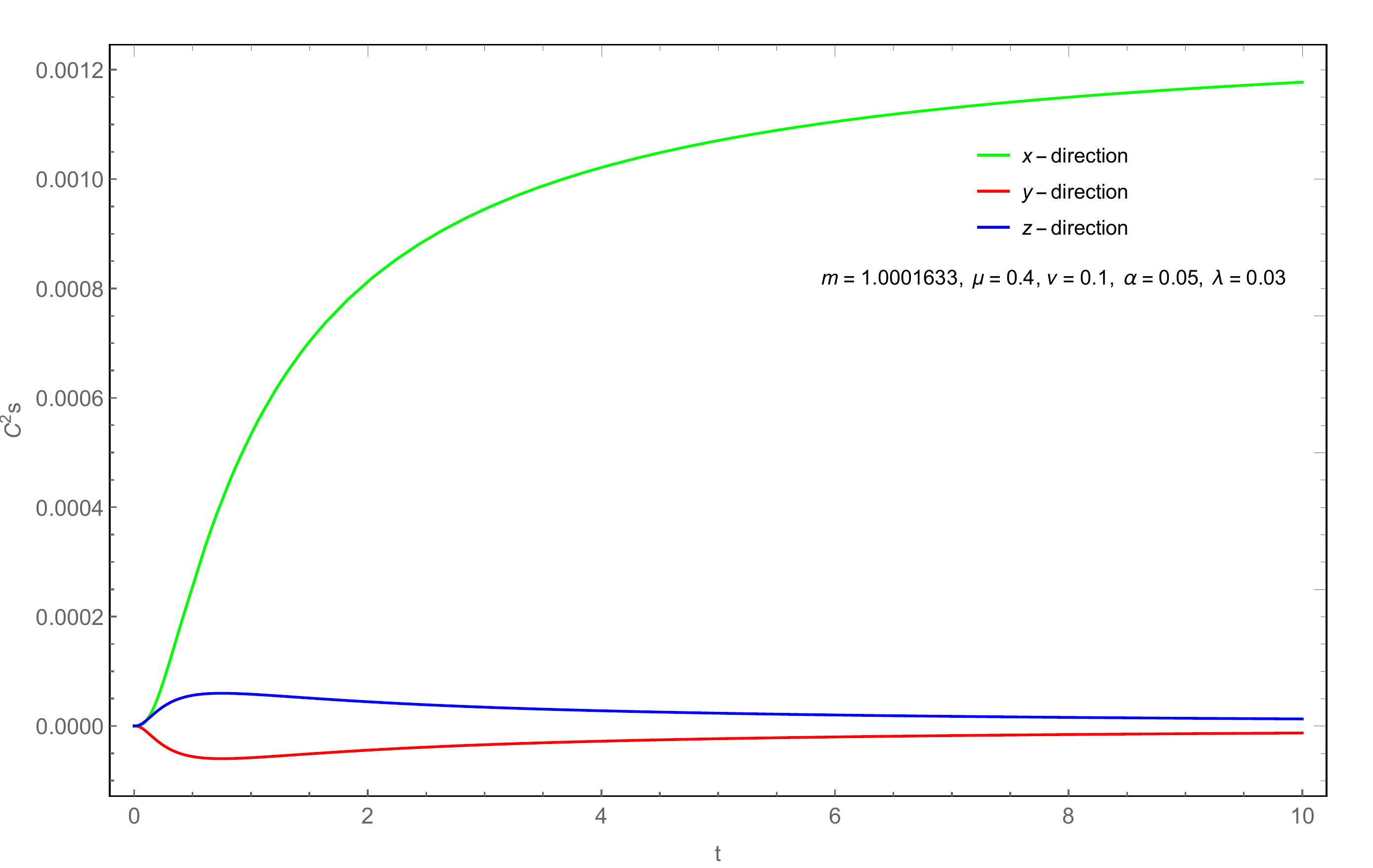}
\caption{Stability analysis for the model with cosmic string and DE in Model III. The unit of the speed of sound $C_s$ is considered as the speed of light in vacuum. The time scale is chosen in the unit of the present age of the Universe i.e at $t=1$, we have the present epoch.}
\label{fig3} 
\end{figure}
\begin{figure}[h!]
\centering
\includegraphics[width=0.8\textwidth]{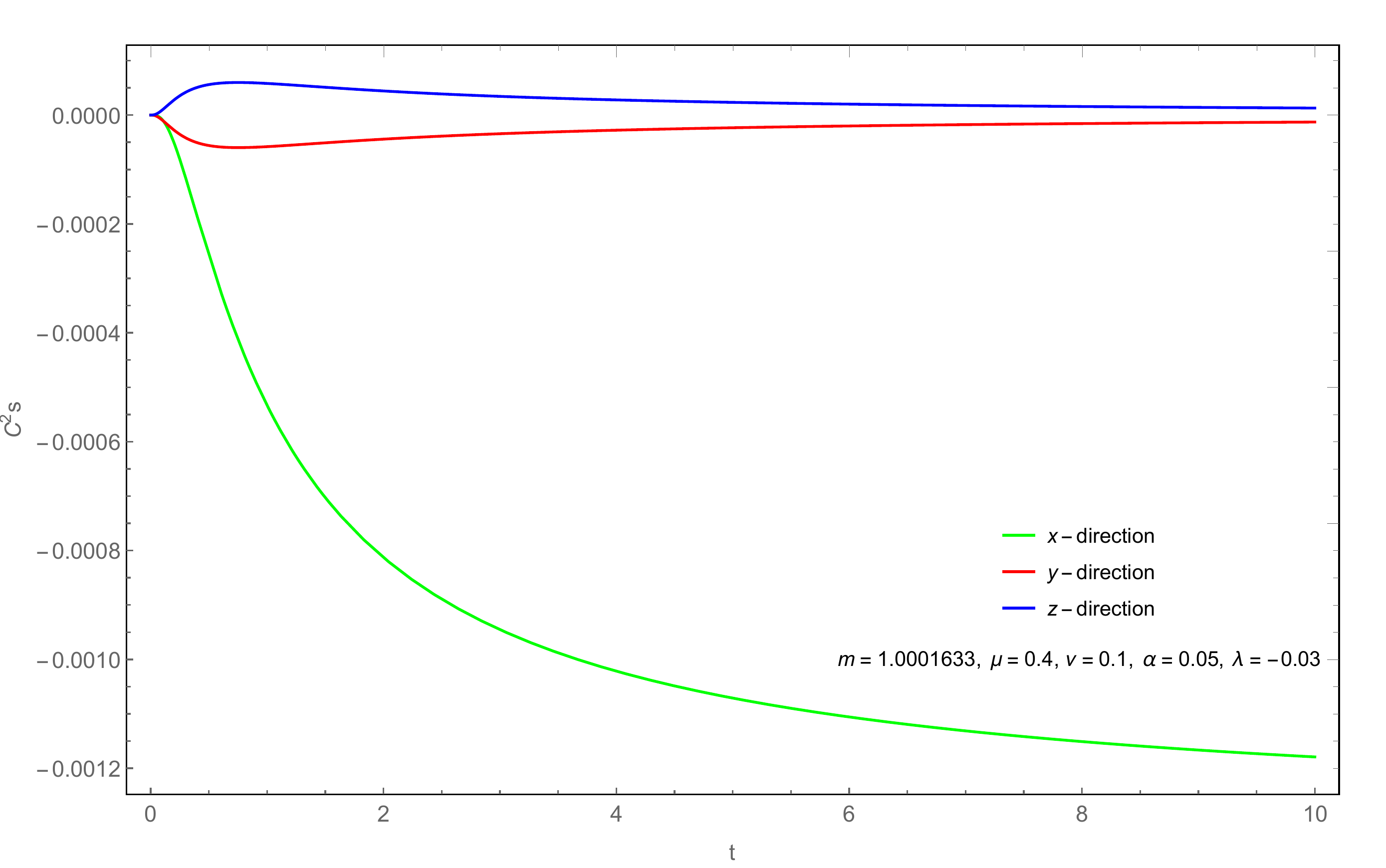}
\caption{Same as FIG. \ref{fig3} but with a negative value of $\lambda$. } 
\end{figure}
The stability of the model with a positive value of $\lambda$ along different spatial directions can be represented in FIG. 3. Since the cosmic strings contribute to the cosmic anisotropy at least in the early phase of cosmic evolution, we expect that, the presence of the cosmic string will affect the stability condition along the $x$-axis to a great extent. With a positive value of $\lambda$, the model is observed to be stable along $x$ and $z$ direction, however remain unstable in $y$-direction. The importance of cosmic string added as a matter field is clearly visible from the graph. The behaviour along $x$-axis is showing a rapid increasing as compared to other two axes with the increase in time for the positive value of the cosmic string. However, for a negative value of the cosmic string, the stability along $x$-direction is disturbed (refer to FIG. 4). 

\subsection{Model IV}

Magnetic field has a great role in controlling the cosmic dynamics at least at early times of cosmic evolution \cite{Maden89, King2007}. It produces large expansion anisotropies during the radiation dominated era \cite{Jacobs69}. Temperature and polarization anisotropies in cosmic microwave background (CMB) radiation is induced by the perturbation due to primordial magnetic field. The effects of large scale magnetic field on CMB radiation can therefore be detected \cite{Kahniashvili2008, Bernui2008}. The origin of cosmic magnetic fields can be attributed to primordial quantum fluctuations and their seeds may be in the range $10^{-18}–10^{-27}$ Gauss or less \cite{Grasso2001, Giovannini2004}. Primordial magnetic field could also have been produced due to cosmological phase transitions \cite{Vachaspati1991}. However, the origin and evolution of large-scale magnetic fields are not yet understood \cite{Durrer2013}. The 21 cm signal, due to the hyperfine transition between 1S singlet and triplet states of the neutral hydrogen atom, is considered as an useful tool to constrain the primordial magnetic field. Recently the Experiment to Detect the Global Epoch of Reionization Signature (EDGES) collaboration observed the absorption signal around 78 MHz in the redshift range $15\lesssim z\lesssim 20$ which indicates that the gas temperature in the intergalactic medium was
cooler than the CMB temperature \cite{Bowman2018}. The EDGES results is nearly two times more than the theoretical prediction based on the $\Lambda$CDM model \cite{Bowman2018}. Basing upon the EDGES results, recently Minoda et al. have constrained the primordial magnetic field at the length scale of 1 Mpc as $B_{1Mpc}\lesssim 10^{-10}$G \cite{Minoda2019}. Using the results of EDGES low-band observation and Absolute Radiometer for Cosmology, Astrophysics and Diffuse Emission (ARCADE 2), Natwariya obtained an upper constraint $B_{1Mpc}\lesssim 53.3 $ pG \cite{Natwariya2020}. In the present work, we wish to consider the presence of magnetic field in addition to the dark energy and investigate its effect on the stability aspects of our model. 

Considering the presence of electromagnetic field along with dark energy \cite{Ray19,Mishra19d}, we may have an additional contribution to the energy momentum tensor as
\begin{equation}
T^M_{ij}=E_{ij}=\frac{1}{4\pi}\left[g^{sp}f_{is}-\frac{1}{4}g_{ij}f_{sp}f^{sp}\right]\label{eq:29}
\end{equation}
where $g_{ij}$ is the gravitational metric and $f_{ij}$ is the electromagnetic field tensor. We assume an infinite electrical conductivity to neglect the effect of electric field on the model which leads to, $f_{14}=f_{24}=f_{34}=0$. Again, quantizing the axis of the magnetic field along the axis of symmetry of longitudinal direction (along the $x$-direction), we got, $f_{12}=f_{13}=0,$ $f_{23}\neq 0$. Thus, the only non-vanishing component of electromagnetic field tensor is $f_{23}$. With the help of Maxwell's equation, the non-vanishing component can be represented as, $f_{23}=-f_{23}= k $, where  $k$ is a constant value, comes from the distribution of magnetic permeability along $x$-direction in the model. Now, incorporating eqn. \eqref{eq:29} in the  set of field eqns. \eqref{eq:9}-\eqref{eq:12}, we obtain 

\begin{eqnarray}
2\dot{H}+4\left(\frac{m^2+m+1}{m^2+2m+1}\right)H^2-\frac{\alpha^2}{\mathcal{R}^2}&=&-\mathcal{M}-(\omega_D+\delta)\rho_D \label{eq:30} \\
\left(\frac{m+3}{m+1}\right)\dot{H}+\left(\frac{m^2+4m+7}{m^2+2m+1}\right)H^2-\frac{\alpha^{2}}{\mathcal{R}^{2}}&=&\mathcal{M}-(\omega_D+\gamma)\rho_D \label{eq:31} \\
\left(\frac{3m+1}{m+1}\right)\dot{H}+\left(\frac{7m^2+4m+1}{m^2+2m+1}\right)H^2-\frac{\alpha^{2}}{\mathcal{R}^{2}}&=&\mathcal{M}-(\omega_D+\eta)\rho_D \label{eq:32} \\
2\left(\frac{m^2+4m+1}{m^2+2m+1}\right)H^2-\frac{3\alpha^{2}}{\mathcal{R}^{2}}&=&-\mathcal{M} +\rho_{D}  \label{eq:33}
\end{eqnarray}

where, $\mathcal{M}=\frac{k^{2}}{8 \pi \mathcal{R}^4 e^{4 \alpha x}}$. Now, eqns. \eqref{eq:30}-\eqref{eq:32} gives the skewness parameters as,
\begin{eqnarray}  
\delta &=& -\frac{2}{3\rho_D}\left[ \left( \frac{m^2-2m+1}{m^2+2m+1} \right) F(R)+\mathcal{M} \right] \\ \label{eq:34}
\gamma &=& \frac{1}{3\rho_D}\left[ \left( \frac{m^2+4m-5}{m^2+2m+1} \right) F(R)+\mathcal{M}\right] \\ \label{eq:35}
\eta &=& -\frac{1}{3\rho_D} \left[\left( \frac{5m^2-4m-1}{m^2+2m+1} \right) F(R)-\mathcal{M}\right]  \label{eq:36}
\end{eqnarray}
The stability function can be calculated using eqns. \eqref{eq:33}-\eqref{eq:36}, and the stability behaviour of the model has been presented in FIGs. 5-7. 
\begin{figure}[h!]
\centering
\includegraphics[width=0.6\textwidth]{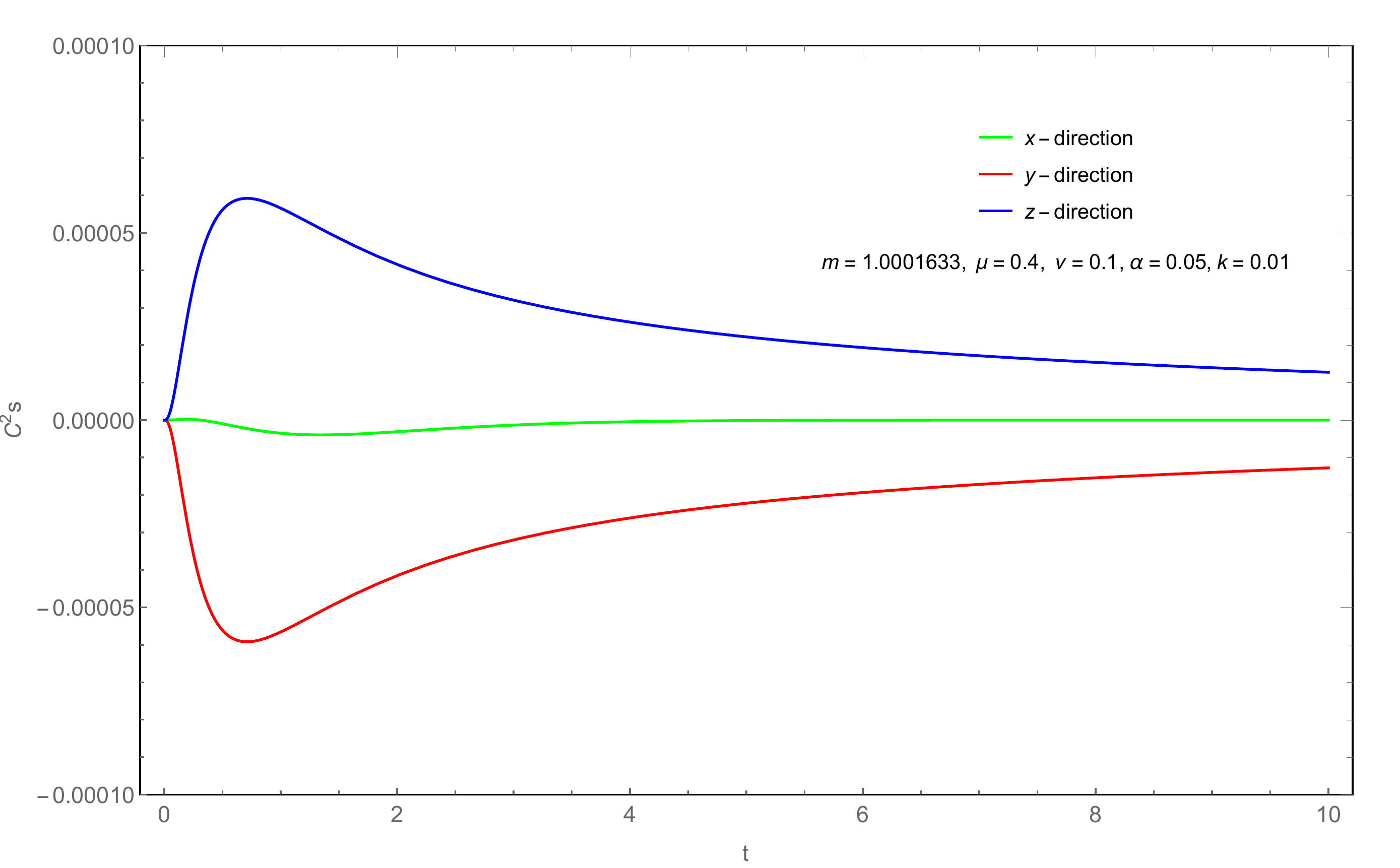}
\caption{Stability analysis for the Model IV with magnetic field and DE for $k=0.01$. The unit of the speed of sound $C_s$ is considered as the speed of light in vacuum. The time scale is chosen in the unit of the present age of the Universe i.e at $t=1$, we have the present epoch.} 
\end{figure}
\begin{figure}[h!]
\centering
\includegraphics[width=0.6\textwidth]{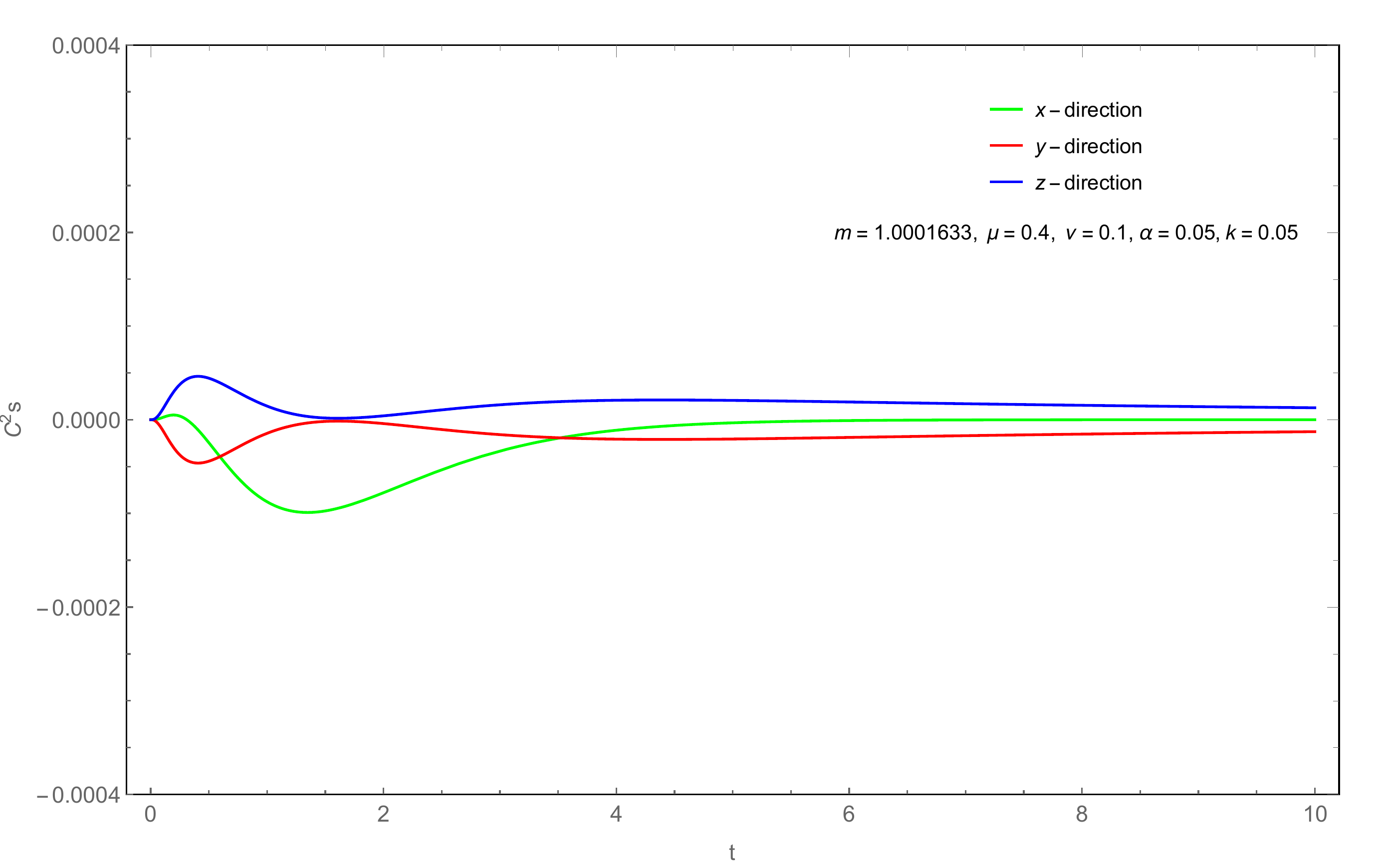}
\caption{Stability analysis for the Model IV with magnetic field and DE for $k=0.05$. The unit of the speed of sound $C_s$ is considered as the speed of light in vacuum. The time scale is chosen in the unit of the present age of the Universe i.e at $t=1$, we have the present epoch.} 
\end{figure}
\begin{figure}[h!]
\centering
\includegraphics[width=0.6\textwidth]{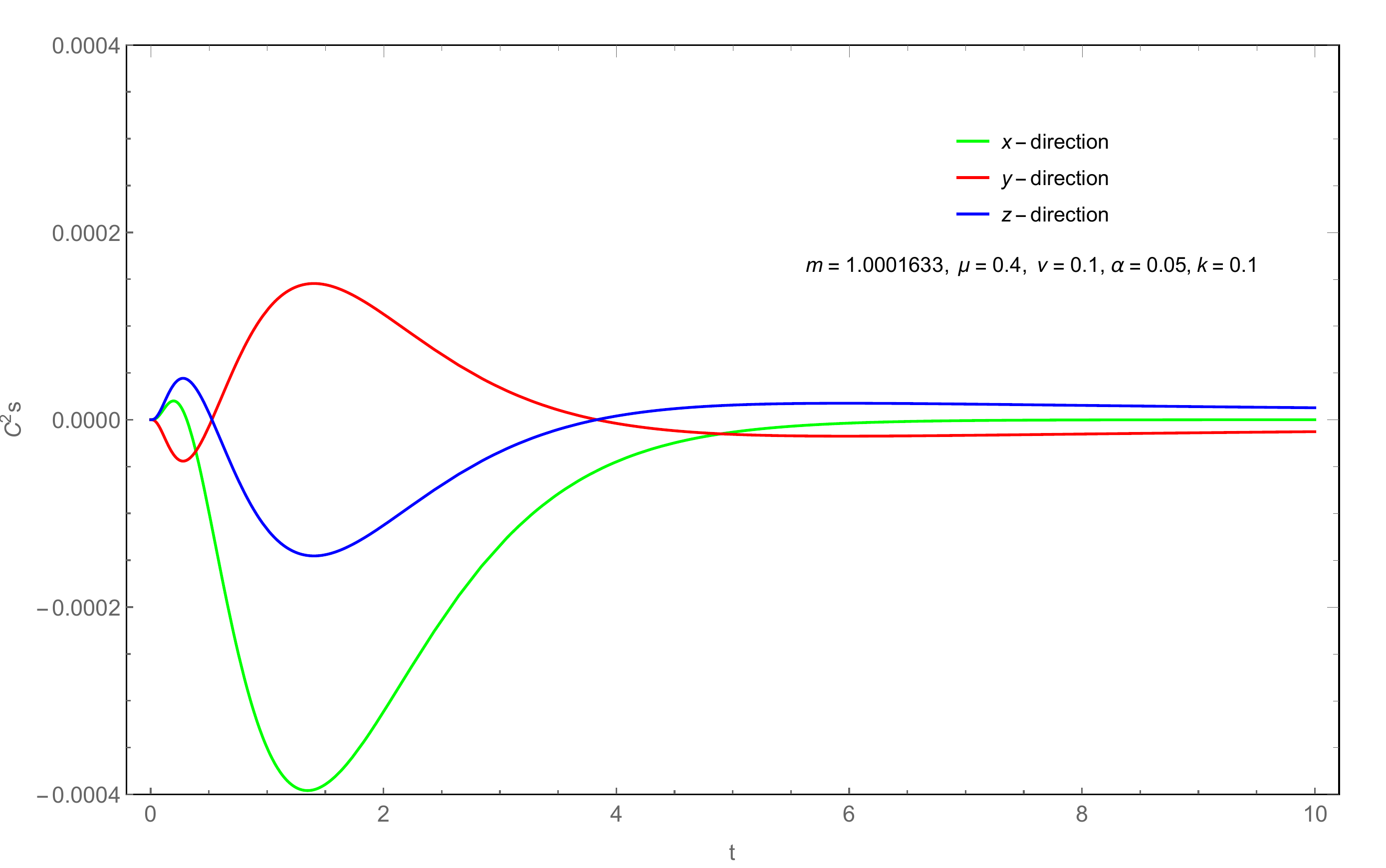}
\caption{Stability analysis for the Model IV with magnetic field and DE for $k=0.1$. The unit of the speed of sound $C_s$ is considered as the speed of light in vacuum. The time scale is chosen in the unit of the present age of the Universe i.e at $t=1$, we have the present epoch.} 
\end{figure}
The addition of magnetic field in the matter field substantially affects  the stability behaviour of the model at least at an initial epoch. Here we have chosen the parameter $k$ to be small representative quantity such as $k=0.01, 0.05$ and $0.1$. The stability aspect is disturbed if we increase the value of  $k$. An increase in the magnetic field particularly affects to the stability along the $x$-axis to a great extent. In the absence of the magnetic field we should expect a behaviour like that in FIG 1 where the stability factor along the $x$-axis is close to zero. However, after switching on the magnetic field, the stability factor along $x$-direction becomes negative. Similarly, the stability aspects along the $y$ and $z$-axes are reversed with an increase in the parameter $k$.
 
\section{Results and Conclusion}
We have analyzed the stability of the cosmological models constructed with the energy momentum tensor as the DE fluid and a combination of DE and matter field. Four cases are presented pertaining to different combination of DE and  matter field in Bianchi V space time. Since the pressure anisotropy is incorporated in the model, the stability analysis has been performed along the spatial directions. The main results of the paper are as follows.
\begin{itemize}
\item In model I and model II along the $x$-direction the parameter determining the mechanical stability of the model vanishes implying that, the models just satisfy the mechanical stability along that direction. This observation may be due to the alignment between the metric potentials and the average scale factor. However with the inclusion of cosmic string in the formalism, model III is observed as stable in this direction.
\item Along $y$ direction, model I and model III is clearly unstable whereas model II showing a signature flipping behaviour, but mostly remains unstable along $y$-direction and stable along $x$ and $z$ direction. At the same time positive and negative value of the cosmic string further affect the behaviour along $x$-direction. The instability along the $y$-direction does not lead to any chaotic behaviour of the model. This has been the result of the presence of an extra negative dark energy pressure along the $y$-direction compared to the usual isotropic pressure. If one considers the magnitude of the departure, then obviously, the quantity defining the mechanical instability along that direction will be a positive quantity.
\item Presence of magnetic field greatly affects the stability condition of the Model IV. For a small magnetic field, the stability aspects of the Model IV and model I are same. The magnetic field disturbs the stability aspect along the $x-$ direction. With an increase in the magnitude of the magnetic field, there occurs a reversal in the trend of stability behaviour along the $y$ and $z$ directions.
\item The effective pressure or the viscous pressure in model II plays some role in the stability behaviour.
\end{itemize} 
We can summarize here that the stability of the models mostly depend on the source of the matter field. If the source of the matter field provides some sort of anisotropy ( such as the one dimensional cosmic strings or magnetic field), the stability condition of the model is substantially affected. In other words, presence of anisotropic sources in the early epoch has a definite role in providing stability to the models.

\section*{Acknowledgement}
BM and SKT thank IUCAA, Pune (India) for hospitality and support during an academic visit where a part of this work is accomplished. BM acknowledges DST, New Delhi, India for providing facilities through DST-FIST lab, Department of Mathematics, where a part of this work was done. The authors are thankful to the anonymous referees for their valuable suggestions and comments for the significant improvement of the paper.

\end{document}